\documentclass[aip, jap, reprint]{revtex4-1}  

\usepackage{color}
\usepackage{graphicx}  
\usepackage{dcolumn}   
\usepackage{bm}        
\usepackage{amsmath}   
\usepackage{amssymb}	

\usepackage[caption=false]{subfig}

\usepackage[colorlinks=true, breaklinks=true, bookmarks=false, linkcolor=blue, urlcolor=blue, citecolor=blue]{hyperref}

\setcitestyle{super}

\begin{document}

\title{Pulsed Laser Deposition of High-Quality Thin Films of the Insulating Ferromagnet EuS}

\author{Qi I. Yang}
\email{qiyang@stanford.edu}
\affiliation{Department of Physics, Stanford University, Stanford, CA 94305}
\affiliation{Stanford Institute for Materials and Energy Sciences, SLAC National Accelerator Laboratory, 2575 Sand Hill Road, Menlo Park, California 94025, USA}
\affiliation{Geballe Laboratory for Advanced Materials, Stanford University, Stanford, CA 94305}

\author{Jinfeng Zhao}
\affiliation{Department of Chemical Engineering and Materials Science, University of California, Davis, CA 95616}

\author{Li Zhang}
\affiliation{Geballe Laboratory for Advanced Materials, Stanford University, Stanford, CA 94305}
\affiliation{Department of Applied Physics, Stanford University, Stanford, CA 94305}

\author{Merav Dolev}
\affiliation{Geballe Laboratory for Advanced Materials, Stanford University, Stanford, CA 94305}
\affiliation{Department of Applied Physics, Stanford University, Stanford, CA 94305}

\author{Alexander D. Fried}
\affiliation{Department of Physics, Stanford University, Stanford, CA 94305}
\affiliation{Geballe Laboratory for Advanced Materials, Stanford University, Stanford, CA 94305}

\author{Ann F. Marshall}
\affiliation{Stanford Nanocharacterization Laboratory, Stanford University, Stanford, CA 94305}

\author{Subhash H. Risbud}
\affiliation{Department of Chemical Engineering and Materials Science, University of California, Davis, CA 95616}

\author{Aharon Kapitulnik}
\affiliation{Department of Physics, Stanford University, Stanford, CA 94305}
\affiliation{Stanford Institute for Materials and Energy Sciences, SLAC National Accelerator Laboratory, 2575 Sand Hill Road, Menlo Park, California 94025, USA}
\affiliation{Geballe Laboratory for Advanced Materials, Stanford University, Stanford, CA 94305}
\affiliation{Department of Applied Physics, Stanford University, Stanford, CA 94305}

\date{\today}

\begin{abstract}
High-quality thin films of the ferromagnetic insulator europium(II) sulfide (EuS) were fabricated by pulsed laser deposition on Al$_2$O$_3$ (0001) and Si (100) substrates. A single orientation was obtained with the [100] planes parallel to the substrates, with atomic-scale smoothness indicates a near-ideal surface topography. The films exhibit uniform ferromagnetism below 15.9~K, with a substantial component of the magnetization perpendicular to the plane of the films. Optimization of the growth condition also yielded truly insulating films with immeasurably large resistance. This combination of magnetic and electric properties open the gate for novel devices that require a true ferromagnetic insulator.
\end{abstract}

\maketitle

Over more than 50 years a wealth of new effects and properties have been discovered in binary lanthanide compounds. In particular, compounds of europium with elements of the sixth group (O,S,Se,Te) exhibit a rock-salt (NaCl)-type crystal structure with ordered magnetic states at low temperatures. As the lattice parameter increases from EuO to EuTe, a ferromagnetic ordered state of moments localized on Eu ions appear in EuO ($T_C\approx69~\mathrm{K}$) and in EuS ($T_C\approx16.7~\mathrm{K}$),\cite{EuO_TC, EuS_Shafer} while EuSe and EuTe show collinear antiferromagnetic ordering with $T_N\approx4.2~\mathrm{K}$, $T_N\approx9.8~\mathrm{K}$ respectively.\cite{EuSe_AF, EuTe_AF} In these chalcogenide compounds, the \textit{S} ground state of Eu$^{2+}$ ions and their simple face centered cubic (FCC) magnetic lattice facilitate testings of the Heisenberg model of ferromagnetism and theories of critical phenomena.\cite{divalent_Eu,  EuS_neighbor_exchange, EuS_critical, EuS_neutron, EuS_spin_wave} A variety of applications were proposed or implemented utilizing these magnetic semiconductors.\cite{EuS_spin_filter, EuS_app1, EuS_spin_filter2} A class of magnetoelectric applications, such as $\pi$-Josephson junctions for quantum qubits~\cite{pi_qubit, pi_junction, Jing} and recently proposed topological magnetoelectric effect associated with the surface state of topological insulators,\cite{QAH_TI_Yu, ISG, MnSe, STS_junction, bilayer} require fabrication of high-quality insulating ferromagnet thin films with robust magnetic properties.

Here we focus on EuS, which is a semiconductor with an indirect energy gap between the 4f$^7$ Eu states and the conduction band minimum at 300~K is 1.65~eV.\cite{EuX_absorption, EuS_band_th1, EuS_band_th2} The lattice parameter of bulk crystals of EuS is $a_0=5.967~\mathrm\AA$, with a ferromagnetic Curie temperature $T_C\approx16.7~\mathrm{K}$. When strained, the lattice constant change is accompanied by a change in Curie temperature, e.g. thin films of EuS grown on KCl show an increase in $T_C$ by as much as 2~K, due to compression induced by differential expansions of the film and substrate.\cite{EuS_PbS} At the same time, very thin films will exhibit slightly lower $T_C$ due to dimensionality reduction. \cite{Keller2002} However, although good electric insulation ($\rho\sim10^{4}~\mathrm{\Omega\cdot}$cm) was obtained in high-quality single crystals, difficulties in material fabrication lead to disorder and unintentional doping, which may drastically reduce the resistivity to as low as $\rho\sim10^{-2}~\mathrm{\Omega\cdot}$cm.\cite{EuS_Shafer, EuS_ntype} Such reduction in resistivity was found to be accompanied by increased Curie temperatures due to interactions between charge carriers and the Eu$^{2+}$ ions.\cite{EuS_TC_doping, EuS_ntype, EuO_TC, EuX_doped_transport} Particularly for thin films, samples fabricated by either pulsed laser deposition (PLD) or molecular beam epitaxy (MBE) were reported to have $T_C$ higher than single crystal values and were suggested to have significant carrier doping.\cite{EuS_MBEa, EuS_MBEb, EuS_PLD1}  In addition, all reported growth methods resulted in samples with multiple crystal orientations,\cite{EuS_PLD1, EuS_PLD2, EuS_MBEb} which might give rise to fractured magnetic domains given the considerable magnetocrystalline anisotropy of EuS.\cite{EuS_anisotropy}  

In this Letter we present results of PLD-grown EuS thin films with significantly improved qualities related to surface topography, magnetic anisotropy, and electrical insulation, all of critical importance for applications involving interfacing the EuS film with another system. Specifically we show characterization results indicating excellent electric insulation, significant and uniform out-of-plane component of the magnetization, a single crystal orientation and a near-ideal surface topography. These improvements should facilitate a series of applications, such as topological insulator--ferromagnet bilayer devices.\cite{bilayer}

For PLD targets, solid disks (approximately $19~\mathrm{mm}$ in diameter and $3~\mathrm{mm}$ thick) were prepared from high-purity (99.95\%) EuS powder by a fast consolidation technique popularly referred to as spark plasma sintering (SPS). This technique uses an electric discharge to activate the surface of the powder particles prior to rapid resistance heating aimed at achieving complete densification. SPS has been effectively used to make solid disk-like targets of a wide range of materials including chalcogenides~\cite{Jinfeng2, Subhash1} and its efficiency in forming clean grain boundaries in polycrystalline targets has been shown for nitrides and refractory high-temperature materials.\cite{Subhash2, Jinfeng1} The target surface was polished with a 800~grit diamond sandpaper before transferring to high vacuum. For final conditioning of the target surface and to deposit EuS thin films, the target was ablated by a 25~ns 248~nm KrF excimer pulsed laser beam in $p=6\times{}10^{-7}~\mathrm{Torr}$ high vacuum at 10~Hz repetition rate. The typical ablation spot size was $2.1\pm0.3~\mathrm{mm^2}$ and the measured fluences were $1.0\pm0.2\mathrm{J\cdot{}cm^{-2}}$. Corundum Al$_2$O$_3$ (0001) and Si (100) substrates were cleaned \textit{ex situ} by solvent sonication prior to transfer to high vacuum. The substrates were heated to $650~^{\circ}\mathrm{C}$ and placed 5~cm away from the target at the center of the plasma plume. The growth rate was estimated to be 1.3~\AA{} per pulse. After each deposition, the substrates were cooled in vacuum to $60~^{\circ}\mathrm{C}$ with a rate slower than $15~^{\circ}\mathrm{C}/\mathrm{min}$.

\begin{figure}[h]%
\subfloat{\label{fig:TEM}}%
\subfloat{\label{fig:AFM}}%
\centering%
\includegraphics[width=\linewidth]{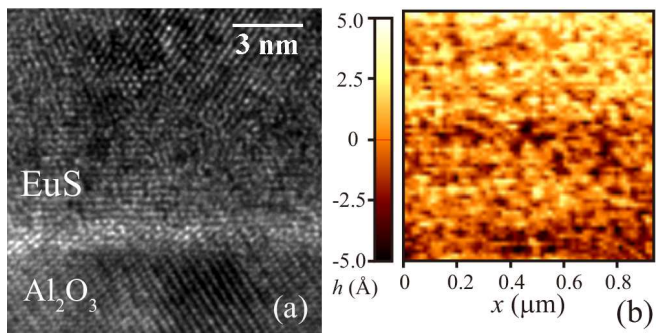}%
\caption{\protect\subref{fig:TEM}~Cross-sectional TEM image of an EuS thin film, showing its interface to the Al$_2$O$_3$ (0001) substrate. \protect\subref{fig:AFM}~AFM image of a $1~\mathrm{\mu{}m}\times{}1~\mathrm{\mu{}m}$ area showing the surface topography of a $20~\mathrm{nm}$ EuS film. The root-mean-square roughness of $\sigma=1.8~\mathrm\AA$ indicates smoothness to the atomic scale.}%
\label{fig:1}%
\end{figure}

The resultant thin films with thicknesses $20~\mathrm{nm}<t<200~\mathrm{nm}$ have a translucent purple color on Al$_2$O$_3$ and are dark green on Si. Fig.~\ref{fig:TEM} shows a transmission electron micrograph (TEM) of a thin film cross-section where the FCC lattice of EuS can be clearly observed. %
The lattice constant is estimated from direct length measurements to be $a=6.0\pm0.2~\mathrm\AA$, consistent with the established results.\cite{EuS_Shafer} Surface topography was measured with an atomic force microscope (AFM). Fig.~\ref{fig:AFM} shows a $1~\mathrm{\mu{}m}\times{}1\mathrm{\mu{}m}$ area on the surface of a 20~nm film on Al$_2$O$_3$. The difference between the minima and maxima in height is roughly twice the lattice constant. The root-mean-square roughness $\sigma=1.8~\mathrm\AA{}$ calculated from a randomly selected line profile indicates near-ideal smoothness. Similar smoothness were obtained on films with thicknesses up to 200~nm deposited on either Al$_2$O$_3$ (0001) or Si (100).

\begin{figure}[h]%
\subfloat{\label{fig:XRD_sap}}%
\subfloat{\label{fig:XRD_Si}}%
\centering%
\includegraphics[width=\linewidth]{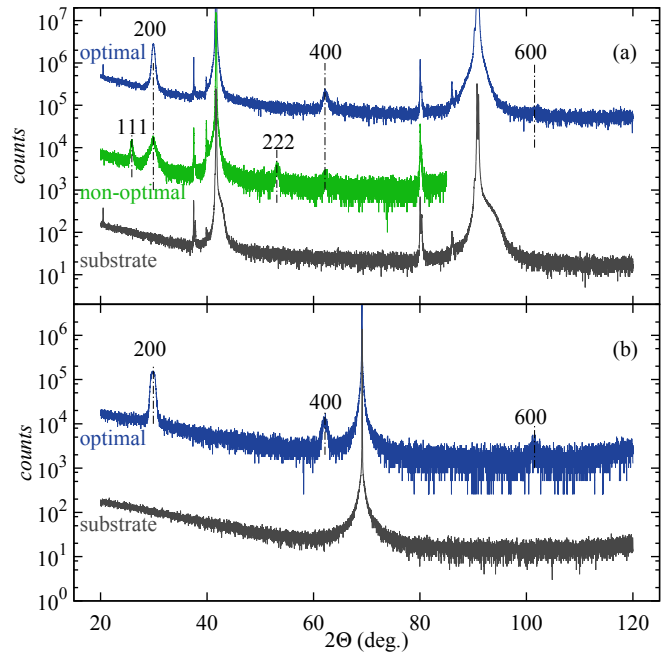}%
\caption{Semi-log X-ray diffraction patterns of 20~nm EuS thin films. \protect\subref{fig:XRD_sap}~On Al$_2$O$_3$ (0001) substrates, optimal growth conditions lead to a single (100) orientation, whereas multiple orientations were observed at non-optimal conditions. Spikes near substrate peaks are due to the K-$\beta$ components of the X-ray source. \protect\subref{fig:XRD_Si}~On Si (100) substrates with native oxides, single (100) orientation was obtained at optimal growth conditions. To distinguish the EuS (400) peak, a monochromator was used in the Si (100) case to eliminate the K-$\beta$ components. (Please see the text for growth conditions.)}%
\label{fig:XRD}%
\end{figure}

Fig.~\ref{fig:XRD} shows the X-ray diffraction patterns on the PLD grown EuS thin films. %
On both Al$_2$O$_3$ (0001) and Si (100) substrates, the optimal conditions described earlier produced samples with a single orientation where the [100] planes are parallel to the substrate surface. On Al$_2$O$_3$ (0001) substrates (fig.~\ref{fig:XRD_sap}), the (200) and (400) reflections are easily identified whereas the (600) reflection is discernible from the background. On Si (100) substrates (fig.~\ref{fig:XRD_Si}), all [100] reflections are clearly observable. For comparison, the diffraction pattern of a non-optimal sample deposited on Al$_2$O$_3$ (0001) at a lower temperature ($T=600~^{\circ}\mathrm{C}$) was plotted in fig.~\ref{fig:XRD_sap}. Reflections from both the [100] and the [111] orientations were observed with comparable weights. Similar multiple orientations were observed in samples deposited at higher-than-optimal temperatures ($T>700~^{\circ}\mathrm{C}$) or higher ambient pressures ($p>2\times{}10^{-6}~\mathrm{Torr}$).

\begin{figure}[h]%
\subfloat{\label{fig:RvT}}%
\subfloat{\label{fig:RvH}}%
\centering%
\includegraphics[width=\linewidth]{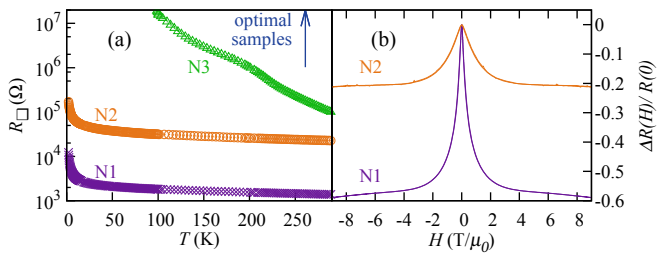}%
\caption{\protect\subref{fig:RvT}~While samples grown at optimal conditions have sheet resistance $R_\Box>20~\mathrm{M\Omega}$ for $2~\mathrm{K}<T<300~\mathrm{K}$, samples grown under non-optimal conditions (N1--N3, with 200~nm thickness) show finite resistance, indicating high carrier densities. \protect\subref{fig:RvH}~These non-optimal samples show negative giant magnetoresistance at $T=2~\mathrm{K}$, similar to that observed in n-type single crystals.}%
\label{fig:transport}
\end{figure}

The resistances of the EuS thin films were measured with a Van der Pauw technique.\cite{VdP1958} When deposited at the optimal conditions on either Al$_2$O$_3$ (0001) or Si (100) substrates, samples with thicknesses $20~\mathrm{nm}<t<200~\mathrm{nm}$ all show sheet resistances $R_\Box>20~\mathrm{M\Omega}$ at temperatures $T > 100~\mathrm{K}$, and immeasurably  high resistance at lower temperatures. This is equivalent to bulk resistivity exceeding  $\rho>400~\mathrm{\Omega\cdot{}cm}$, consistent to values obtained on high-purity single crystals.\cite{EuS_Shafer} In contrast, films deposited at non-optimal conditions show sheet resistances as low as $R_\Box\sim~\mathrm{k\Omega}$ (fig.~\ref{fig:RvT}), %
which corresponds to a bulk resistivity of $\rho\sim10^{-2}\mathrm{\Omega\cdot{}cm}$, consistent with the conductive r\'egime in doped single crystals.\cite{EuS_ntype} 
Different from n-type single crystals results, where resistance anomalies were observed near $T_C$ and attributed to change in carrier concentrations,\cite{EuS_ntype, EuX_doped_transport} monotonic increases in resistance were observed towards low temperatures in thin films. Such difference could be resulted from different natures of dopants or due to the effects of reduced dimensionality.\cite{2D_conduction} Similar to n-type doped single crystals, negative giant magnetoresistance was observed in conducting samples at low temperatures (fig.~\ref{fig:RvH}).

\begin{figure}[h]%
\subfloat{\label{fig:MvH_z}}%
\subfloat{\label{fig:MvH_x}}%
\subfloat{\label{fig:MvT_z}}%
\subfloat{\label{fig:Sagnac}}%
\centering%
\includegraphics[width=\linewidth]{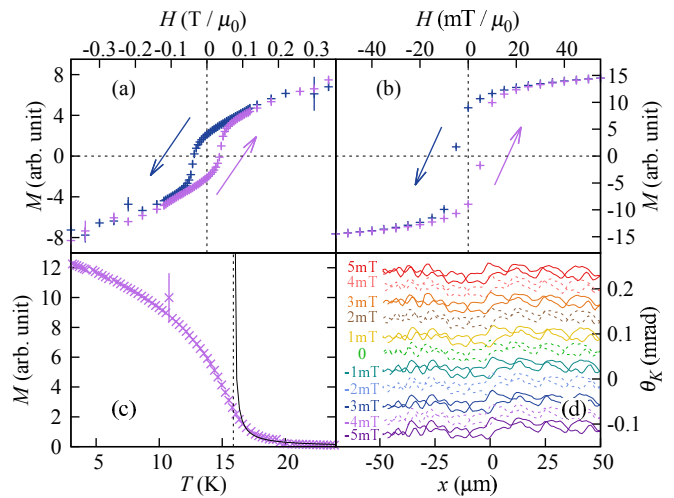}%
\caption{Magnetization of a 20~nm EuS film on Al$_2$O$_3$ (0001), \protect\subref{fig:MvH_z}~in perpendicular fields, \protect\subref{fig:MvH_x}~in parallel fields, and \protect\subref{fig:MvT_z}~its temperature dependence, plotted in the same arbitrary unit with a linear paramagnetic component of the substrate subtracted. The fitting to the Curie-Weiss Law indicates a low Curie temperature $T_C=15.9~\mathrm{K}$. \protect\subref{fig:Sagnac}~Kerr effect measured with a scanning Sagnac interferometer at $T=10~\mathrm{K}$, showing uniform magnetization.}%
\label{fig:magnetic}%
\end{figure}

Magnetizations of the thin films were measured in a superconducting quantum interference device (SQUID) magnetometer down to $T=2~\mathrm{K}$. A significant perpendicular component of the magnetization was observed (fig.~\ref{fig:MvH_z}), whereas the easy axes are in the sample plane (fig.~\ref{fig:MvH_x}). %
While the coercive field of perpendicular magnetization may vary within the same order of magnitude for film thicknesses between 20~nm and 200~nm, the general hysteresis features are similar for all our samples on either Al$_2$O$_3$ (0001) or Si (100). By fitting to the Curie-Weiss law in the paramagnetic regime (fig.~\ref{fig:MvT_z}), an upper limit of the Curie temperature of an optimal 20~nm thin film on Al$_2$O$_3$ (0001) was estimated to be $T_C=15.9~\mathrm{K}$.~\cite{Eu_mag_compounds} While a low $T_C$ is expected for samples with diminishing carrier densities,~\cite{EuS_ntype, EuO_TC, EuX_doped_transport} we note that the thin film $T_C$ is lower than the minimal single crystal value $T_C\approx16.5~\mathrm{K}$. This is likely a combined effect of poly-crystallinity, approaching the two-dimensional limit and lattice strains.~\cite{TC_thickness1, TC_thickness2, TC_strain}

To test for homogeneity of the magnetism in our films we used a scanning Sagnac interferometer.~\cite{futureSagnac} This device is based on a zero-area loop Sagnac interferometer as was first demonstrated by Xia {\it et al.},~\cite{xia:062508} and can measure the Polar Kerr angle upon reflection from the film with shot-noise limited sensitivity at low power. Our scanning device is operated at a wavelength of 820~nm, and has a spatial resolution of 0.9~$\mu$m. Fig.~\ref{fig:Sagnac} shows several line scans of length~100 $\mu$m, taken at a temperature of 10~K and at low magnetic fields,  showing a very uniform Kerr response. This set of line scans also agrees with the coercive field found in the SQUID magnetometry measurements.

While the experimental setup for PLD is relatively simple, it is well known that complex and non-equilibrium mechanisms are involved in both laser ablation of the target and plume-substrate interactions. Here we present a tentative discussion on the growth process. At the optimal growth conditions, the resultant atomically smooth topography seems to suggest the Frank--van der Merwe mode of nucleation. The absence of microstructures, which indicates sufficient reduction of partial evaporation (``splashing''),~\cite{PLD_book} may have been in part due to the effective densification with the SPS technique and appropriate target surface treatment. While EuS solid is stable up to $2300~^{\circ}\mathrm{C}$ in vacuum,\cite{EuS_phase_diagram} we found that the film quality is sensitive to relatively small deviations ($\Delta{}T=\pm50~^{\circ}\mathrm{C}$) from the optimal substrate temperature $T=650~^{\circ}\mathrm{C}$. Lowering the substrate temperature or increasing the ambient pressure are known to increase the cooling rate of adatoms. For compounds with large differences in constituent elements' vapor pressures such as EuS, ($p(\mathrm{S})/p(\mathrm{Eu})>10^4$ at $T=650~^{\circ}\mathrm{C}$,~\cite{elements}) such effects worsen both stoichiometry and structure.~\cite{Metev1988} In addition, the detrimental effects of changing temperature at either directions may be related to the nearby eutectic point at $750~^{\circ}\mathrm{C}$ and the EuS$_2$ phase below $575~^{\circ}\mathrm{C}${ },~\cite{EuS_phase_diagram} which may provide transient states that facilitate structural distortions or an Eu-rich stoichiometry. In either case, the distortion in stoichiometry would likely result in unintended doping.

To summarize, high-quality thin films of EuS were fabricated by pulsed laser deposition. Single (100) orientation and atomic-scale surface smoothness ($\sigma=1.8~\mathrm\AA{}$) were obtained. Samples deposited at optimal conditions are highly insulating with sheet resistance $R_\Box>20~\mathrm{M\Omega}$ for thickness $20~\mathrm{nm}<t<200~\mathrm{nm}$. Significant magnetizations were observed at the out-of-plane direction, showing hysteresis and homogeneous spatial distribution. These properties are crucial for various magnetoelectric applications. For comparison, we demonstrated that samples deposited at non-optimal conditions show low resistance and negative giant magnetoresistance, indicating unintended doping, possibly due to distorted composition stoichiometry.

\begin{acknowledgments}
We would like to thank Robert Hammond, Min Liu, Di Lu, Alexander Palevski, Elizabeth Schemm, Arturas Vailionis and Philip Wu for help and discussions. This work is supported by DARPA, MesoDynamic Architecture Program (MESO) through the contract number N66001-11-1-4105,  by FENA, and by a seed grant from DOE for the study of TI. A series of open-source software have been utilized during data analysis.~\cite{gsl, gnuplot}
\end{acknowledgments}

\bibliography{EuSPLD}

\begin{thebibliography}{51}%
\makeatletter
\providecommand \@ifxundefined [1]{%
 \@ifx{#1\undefined}
}%
\providecommand \@ifnum [1]{%
 \ifnum #1\expandafter \@firstoftwo
 \else \expandafter \@secondoftwo
 \fi
}%
\providecommand \@ifx [1]{%
 \ifx #1\expandafter \@firstoftwo
 \else \expandafter \@secondoftwo
 \fi
}%
\providecommand \natexlab [1]{#1}%
\providecommand \enquote  [1]{``#1''}%
\providecommand \bibnamefont  [1]{#1}%
\providecommand \bibfnamefont [1]{#1}%
\providecommand \citenamefont [1]{#1}%
\providecommand \href@noop [0]{\@secondoftwo}%
\providecommand \href [0]{\begingroup \@sanitize@url \@href}%
\providecommand \@href[1]{\@@startlink{#1}\@@href}%
\providecommand \@@href[1]{\endgroup#1\@@endlink}%
\providecommand \@sanitize@url [0]{\catcode `\\12\catcode `\$12\catcode
  `\&12\catcode `\#12\catcode `\^12\catcode `\_12\catcode `\%12\relax}%
\providecommand \@@startlink[1]{}%
\providecommand \@@endlink[0]{}%
\providecommand \url  [0]{\begingroup\@sanitize@url \@url }%
\providecommand \@url [1]{\endgroup\@href {#1}{\urlprefix }}%
\providecommand \urlprefix  [0]{URL }%
\providecommand \Eprint [0]{\href }%
\providecommand \doibase [0]{http://dx.doi.org/}%
\providecommand \selectlanguage [0]{\@gobble}%
\providecommand \bibinfo  [0]{\@secondoftwo}%
\providecommand \bibfield  [0]{\@secondoftwo}%
\providecommand \translation [1]{[#1]}%
\providecommand \BibitemOpen [0]{}%
\providecommand \bibitemStop [0]{}%
\providecommand \bibitemNoStop [0]{.\EOS\space}%
\providecommand \EOS [0]{\spacefactor3000\relax}%
\providecommand \BibitemShut  [1]{\csname bibitem#1\endcsname}%
\let\auto@bib@innerbib\@empty
\bibitem [{\citenamefont {Shafer}\ and\ \citenamefont
  {McGuire}(1968)}]{EuO_TC}%
  \BibitemOpen
  \bibfield  {author} {\bibinfo {author} {\bibfnamefont {M.~W.}\ \bibnamefont
  {Shafer}}\ and\ \bibinfo {author} {\bibfnamefont {T.~R.}\ \bibnamefont
  {McGuire}},\ }\href {\doibase 10.1063/1.2163533} {\bibfield  {journal}
  {\bibinfo  {journal} {Journal of Applied Physics}\ }\textbf {\bibinfo
  {volume} {39}},\ \bibinfo {pages} {588} (\bibinfo {year} {1968})}\BibitemShut
  {NoStop}%
\bibitem [{\citenamefont {Shafer}(1972)}]{EuS_Shafer}%
  \BibitemOpen
  \bibfield  {author} {\bibinfo {author} {\bibfnamefont {M.}~\bibnamefont
  {Shafer}},\ }\href {\doibase 10.1016/0025-5408(72)90258-9} {\bibfield
  {journal} {\bibinfo  {journal} {Materials Research Bulletin}\ }\textbf
  {\bibinfo {volume} {7}},\ \bibinfo {pages} {603 } (\bibinfo {year}
  {1972})}\BibitemShut {NoStop}%
\bibitem [{\citenamefont {Fischer}\ \emph {et~al.}(1969)\citenamefont
  {Fischer}, \citenamefont {Hälg}, \citenamefont {Wartburg}, \citenamefont
  {Schwob},\ and\ \citenamefont {Vogt}}]{EuSe_AF}%
  \BibitemOpen
  \bibfield  {author} {\bibinfo {author} {\bibfnamefont {P.}~\bibnamefont
  {Fischer}}, \bibinfo {author} {\bibfnamefont {W.}~\bibnamefont {Hälg}},
  \bibinfo {author} {\bibfnamefont {W.}~\bibnamefont {Wartburg}}, \bibinfo
  {author} {\bibfnamefont {P.}~\bibnamefont {Schwob}}, \ and\ \bibinfo {author}
  {\bibfnamefont {O.}~\bibnamefont {Vogt}},\ }\href {\doibase
  10.1007/BF02422568} {\bibfield  {journal} {\bibinfo  {journal} {Physik der
  kondensierten Materie}\ }\textbf {\bibinfo {volume} {9}},\ \bibinfo {pages}
  {249} (\bibinfo {year} {1969})}\BibitemShut {NoStop}%
\bibitem [{\citenamefont {Busch}\ \emph {et~al.}(1964)\citenamefont {Busch},
  \citenamefont {Junod}, \citenamefont {Morris}, \citenamefont {Muheim},\ and\
  \citenamefont {Stutius}}]{EuTe_AF}%
  \BibitemOpen
  \bibfield  {author} {\bibinfo {author} {\bibfnamefont {G.}~\bibnamefont
  {Busch}}, \bibinfo {author} {\bibfnamefont {P.}~\bibnamefont {Junod}},
  \bibinfo {author} {\bibfnamefont {R.}~\bibnamefont {Morris}}, \bibinfo
  {author} {\bibfnamefont {J.}~\bibnamefont {Muheim}}, \ and\ \bibinfo {author}
  {\bibfnamefont {W.}~\bibnamefont {Stutius}},\ }\href {\doibase
  http://dx.doi.org/10.1016/0031-9163(64)90230-6} {\bibfield  {journal}
  {\bibinfo  {journal} {Physics Letters}\ }\textbf {\bibinfo {volume} {11}},\
  \bibinfo {pages} {9 } (\bibinfo {year} {1964})}\BibitemShut {NoStop}%
\bibitem [{\citenamefont {McGuire}\ \emph {et~al.}(1963)\citenamefont
  {McGuire}, \citenamefont {Argyle}, \citenamefont {Shafer},\ and\
  \citenamefont {Smart}}]{divalent_Eu}%
  \BibitemOpen
  \bibfield  {author} {\bibinfo {author} {\bibfnamefont {T.~R.}\ \bibnamefont
  {McGuire}}, \bibinfo {author} {\bibfnamefont {B.~E.}\ \bibnamefont {Argyle}},
  \bibinfo {author} {\bibfnamefont {M.~W.}\ \bibnamefont {Shafer}}, \ and\
  \bibinfo {author} {\bibfnamefont {J.~S.}\ \bibnamefont {Smart}},\ }\href
  {\doibase 10.1063/1.1729501} {\bibfield  {journal} {\bibinfo  {journal}
  {Journal of Applied Physics}\ }\textbf {\bibinfo {volume} {34}},\ \bibinfo
  {pages} {1345} (\bibinfo {year} {1963})}\BibitemShut {NoStop}%
\bibitem [{\citenamefont {Wojtowicz}(1964)}]{EuS_neighbor_exchange}%
  \BibitemOpen
  \bibfield  {author} {\bibinfo {author} {\bibfnamefont {P.~J.}\ \bibnamefont
  {Wojtowicz}},\ }\href {\doibase 10.1063/1.1713571} {\bibfield  {journal}
  {\bibinfo  {journal} {Journal of Applied Physics}\ }\textbf {\bibinfo
  {volume} {35}},\ \bibinfo {pages} {991} (\bibinfo {year} {1964})}\BibitemShut
  {NoStop}%
\bibitem [{\citenamefont {Boni}\ \emph {et~al.}(1987)\citenamefont {Boni},
  \citenamefont {Shirane}, \citenamefont {Bohn},\ and\ \citenamefont
  {Zinn}}]{EuS_critical}%
  \BibitemOpen
  \bibfield  {author} {\bibinfo {author} {\bibfnamefont {P.}~\bibnamefont
  {Boni}}, \bibinfo {author} {\bibfnamefont {G.}~\bibnamefont {Shirane}},
  \bibinfo {author} {\bibfnamefont {H.~G.}\ \bibnamefont {Bohn}}, \ and\
  \bibinfo {author} {\bibfnamefont {W.}~\bibnamefont {Zinn}},\ }\href {\doibase
  10.1063/1.338786} {\bibfield  {journal} {\bibinfo  {journal} {Journal of
  Applied Physics}\ }\textbf {\bibinfo {volume} {61}},\ \bibinfo {pages} {3397}
  (\bibinfo {year} {1987})}\BibitemShut {NoStop}%
\bibitem [{\citenamefont {Bohn}, \citenamefont {Kollmar},\ and\ \citenamefont
  {Zinn}(1984)}]{EuS_neutron}%
  \BibitemOpen
  \bibfield  {author} {\bibinfo {author} {\bibfnamefont {H.~G.}\ \bibnamefont
  {Bohn}}, \bibinfo {author} {\bibfnamefont {A.}~\bibnamefont {Kollmar}}, \
  and\ \bibinfo {author} {\bibfnamefont {W.}~\bibnamefont {Zinn}},\ }\href
  {\doibase 10.1103/PhysRevB.30.6504} {\bibfield  {journal} {\bibinfo
  {journal} {Phys. Rev. B}\ }\textbf {\bibinfo {volume} {30}},\ \bibinfo
  {pages} {6504} (\bibinfo {year} {1984})}\BibitemShut {NoStop}%
\bibitem [{\citenamefont {Charap}\ and\ \citenamefont
  {Boyd}(1964)}]{EuS_spin_wave}%
  \BibitemOpen
  \bibfield  {author} {\bibinfo {author} {\bibfnamefont {S.~H.}\ \bibnamefont
  {Charap}}\ and\ \bibinfo {author} {\bibfnamefont {E.~L.}\ \bibnamefont
  {Boyd}},\ }\href {\doibase 10.1103/PhysRev.133.A811} {\bibfield  {journal}
  {\bibinfo  {journal} {Phys. Rev.}\ }\textbf {\bibinfo {volume} {133}},\
  \bibinfo {pages} {A811} (\bibinfo {year} {1964})}\BibitemShut {NoStop}%
\bibitem [{\citenamefont {Esaki}, \citenamefont {Stiles},\ and\ \citenamefont
  {Molnar}(1967)}]{EuS_spin_filter}%
  \BibitemOpen
  \bibfield  {author} {\bibinfo {author} {\bibfnamefont {L.}~\bibnamefont
  {Esaki}}, \bibinfo {author} {\bibfnamefont {P.~J.}\ \bibnamefont {Stiles}}, \
  and\ \bibinfo {author} {\bibfnamefont {S.~v.}\ \bibnamefont {Molnar}},\
  }\href {\doibase 10.1103/PhysRevLett.19.852} {\bibfield  {journal} {\bibinfo
  {journal} {Phys. Rev. Lett.}\ }\textbf {\bibinfo {volume} {19}},\ \bibinfo
  {pages} {852} (\bibinfo {year} {1967})}\BibitemShut {NoStop}%
\bibitem [{\citenamefont {Freeman}(1995)}]{EuS_app1}%
  \BibitemOpen
  \bibfield  {author} {\bibinfo {author} {\bibfnamefont {M.~R.}\ \bibnamefont
  {Freeman}},\ }\href@noop {} {\enquote {\bibinfo {title} {Fiber optic probe
  with a magneto-optic film on an end surface for detecting a current in an
  integrated circuit},}\ } (\bibinfo {year} {1995}),\ \bibinfo {note} {{US}
  Patent 5,451,863}\BibitemShut {NoStop}%
\bibitem [{\citenamefont {M\"uller}, \citenamefont {Luysberg},\ and\
  \citenamefont {Schneider}(2011)}]{EuS_spin_filter2}%
  \BibitemOpen
  \bibfield  {author} {\bibinfo {author} {\bibfnamefont {M.}~\bibnamefont
  {M\"uller}}, \bibinfo {author} {\bibfnamefont {M.}~\bibnamefont {Luysberg}},
  \ and\ \bibinfo {author} {\bibfnamefont {C.~M.}\ \bibnamefont {Schneider}},\
  }\href {\doibase 10.1063/1.3572016} {\bibfield  {journal} {\bibinfo
  {journal} {Applied Physics Letters}\ }\textbf {\bibinfo {volume} {98}},\
  \bibinfo {eid} {142503} (\bibinfo {year} {2011})}\BibitemShut {NoStop}%
\bibitem [{\citenamefont {Yamashita}\ \emph {et~al.}(2005)\citenamefont
  {Yamashita}, \citenamefont {Tanikawa}, \citenamefont {Takahashi},\ and\
  \citenamefont {Maekawa}}]{pi_qubit}%
  \BibitemOpen
  \bibfield  {author} {\bibinfo {author} {\bibfnamefont {T.}~\bibnamefont
  {Yamashita}}, \bibinfo {author} {\bibfnamefont {K.}~\bibnamefont {Tanikawa}},
  \bibinfo {author} {\bibfnamefont {S.}~\bibnamefont {Takahashi}}, \ and\
  \bibinfo {author} {\bibfnamefont {S.}~\bibnamefont {Maekawa}},\ }\href
  {\doibase 10.1103/PhysRevLett.95.097001} {\bibfield  {journal} {\bibinfo
  {journal} {Phys. Rev. Lett.}\ }\textbf {\bibinfo {volume} {95}},\ \bibinfo
  {pages} {097001} (\bibinfo {year} {2005})}\BibitemShut {NoStop}%
\bibitem [{\citenamefont {Bulaevskii}, \citenamefont {Kuzii},\ and\
  \citenamefont {Sobyanin}(1977)}]{pi_junction}%
  \BibitemOpen
  \bibfield  {author} {\bibinfo {author} {\bibfnamefont {L.}~\bibnamefont
  {Bulaevskii}}, \bibinfo {author} {\bibfnamefont {V.}~\bibnamefont {Kuzii}}, \
  and\ \bibinfo {author} {\bibfnamefont {A.}~\bibnamefont {Sobyanin}},\
  }\href@noop {} {\bibfield  {journal} {\bibinfo  {journal} {JETP lett.}\
  }\textbf {\bibinfo {volume} {25}},\ \bibinfo {pages} {290} (\bibinfo {year}
  {1977})}\BibitemShut {NoStop}%
\bibitem [{\citenamefont {Xia}\ \emph {et~al.}(2009)\citenamefont {Xia},
  \citenamefont {Shelukhin}, \citenamefont {Karpovski}, \citenamefont
  {Kapitulnik},\ and\ \citenamefont {Palevski}}]{Jing}%
  \BibitemOpen
  \bibfield  {author} {\bibinfo {author} {\bibfnamefont {J.}~\bibnamefont
  {Xia}}, \bibinfo {author} {\bibfnamefont {V.}~\bibnamefont {Shelukhin}},
  \bibinfo {author} {\bibfnamefont {M.}~\bibnamefont {Karpovski}}, \bibinfo
  {author} {\bibfnamefont {A.}~\bibnamefont {Kapitulnik}}, \ and\ \bibinfo
  {author} {\bibfnamefont {A.}~\bibnamefont {Palevski}},\ }\href {\doibase
  10.1103/PhysRevLett.102.087004} {\bibfield  {journal} {\bibinfo  {journal}
  {Phys. Rev. Lett.}\ }\textbf {\bibinfo {volume} {102}},\ \bibinfo {pages}
  {087004} (\bibinfo {year} {2009})}\BibitemShut {NoStop}%
\bibitem [{\citenamefont {Yu}\ \emph {et~al.}(2010)\citenamefont {Yu},
  \citenamefont {Zhang}, \citenamefont {Zhang}, \citenamefont {Zhang},
  \citenamefont {Dai},\ and\ \citenamefont {Fang}}]{QAH_TI_Yu}%
  \BibitemOpen
  \bibfield  {author} {\bibinfo {author} {\bibfnamefont {R.}~\bibnamefont
  {Yu}}, \bibinfo {author} {\bibfnamefont {W.}~\bibnamefont {Zhang}}, \bibinfo
  {author} {\bibfnamefont {H.-J.}\ \bibnamefont {Zhang}}, \bibinfo {author}
  {\bibfnamefont {S.-C.}\ \bibnamefont {Zhang}}, \bibinfo {author}
  {\bibfnamefont {X.}~\bibnamefont {Dai}}, \ and\ \bibinfo {author}
  {\bibfnamefont {Z.}~\bibnamefont {Fang}},\ }\href {\doibase
  10.1126/science.1187485} {\bibfield  {journal} {\bibinfo  {journal}
  {Science}\ }\textbf {\bibinfo {volume} {329}},\ \bibinfo {pages} {61}
  (\bibinfo {year} {2010})}\BibitemShut {NoStop}%
\bibitem [{\citenamefont {Garate}\ and\ \citenamefont {Franz}(2010)}]{ISG}%
  \BibitemOpen
  \bibfield  {author} {\bibinfo {author} {\bibfnamefont {I.}~\bibnamefont
  {Garate}}\ and\ \bibinfo {author} {\bibfnamefont {M.}~\bibnamefont {Franz}},\
  }\href {\doibase 10.1103/PhysRevLett.104.146802} {\bibfield  {journal}
  {\bibinfo  {journal} {Phys. Rev. Lett.}\ }\textbf {\bibinfo {volume} {104}},\
  \bibinfo {pages} {146802} (\bibinfo {year} {2010})}\BibitemShut {NoStop}%
\bibitem [{\citenamefont {Luo}\ and\ \citenamefont {Qi}(2013)}]{MnSe}%
  \BibitemOpen
  \bibfield  {author} {\bibinfo {author} {\bibfnamefont {W.}~\bibnamefont
  {Luo}}\ and\ \bibinfo {author} {\bibfnamefont {X.-L.}\ \bibnamefont {Qi}},\
  }\href {\doibase 10.1103/PhysRevB.87.085431} {\bibfield  {journal} {\bibinfo
  {journal} {Phys. Rev. B}\ }\textbf {\bibinfo {volume} {87}},\ \bibinfo
  {pages} {085431} (\bibinfo {year} {2013})}\BibitemShut {NoStop}%
\bibitem [{\citenamefont {Snelder}\ \emph {et~al.}(2013)\citenamefont
  {Snelder}, \citenamefont {Veldhorst}, \citenamefont {Golubov},\ and\
  \citenamefont {Brinkman}}]{STS_junction}%
  \BibitemOpen
  \bibfield  {author} {\bibinfo {author} {\bibfnamefont {M.}~\bibnamefont
  {Snelder}}, \bibinfo {author} {\bibfnamefont {M.}~\bibnamefont {Veldhorst}},
  \bibinfo {author} {\bibfnamefont {A.~A.}\ \bibnamefont {Golubov}}, \ and\
  \bibinfo {author} {\bibfnamefont {A.}~\bibnamefont {Brinkman}},\ }\href
  {\doibase 10.1103/PhysRevB.87.104507} {\bibfield  {journal} {\bibinfo
  {journal} {Phys. Rev. B}\ }\textbf {\bibinfo {volume} {87}},\ \bibinfo
  {pages} {104507} (\bibinfo {year} {2013})}\BibitemShut {NoStop}%
\bibitem [{\citenamefont {Yang}\ \emph {et~al.}(2013)\citenamefont {Yang},
  \citenamefont {Dolev}, \citenamefont {Zhang}, \citenamefont {Zhao},
  \citenamefont {Fried}, \citenamefont {Schemm}, \citenamefont {Liu},
  \citenamefont {Palevski}, \citenamefont {Marshall}, \citenamefont {Risbud},\
  and\ \citenamefont {Kapitulnik}}]{bilayer}%
  \BibitemOpen
  \bibfield  {author} {\bibinfo {author} {\bibfnamefont {Q.~I.}\ \bibnamefont
  {Yang}}, \bibinfo {author} {\bibfnamefont {M.}~\bibnamefont {Dolev}},
  \bibinfo {author} {\bibfnamefont {L.}~\bibnamefont {Zhang}}, \bibinfo
  {author} {\bibfnamefont {J.}~\bibnamefont {Zhao}}, \bibinfo {author}
  {\bibfnamefont {A.~D.}\ \bibnamefont {Fried}}, \bibinfo {author}
  {\bibfnamefont {E.}~\bibnamefont {Schemm}}, \bibinfo {author} {\bibfnamefont
  {M.}~\bibnamefont {Liu}}, \bibinfo {author} {\bibfnamefont {A.}~\bibnamefont
  {Palevski}}, \bibinfo {author} {\bibfnamefont {A.~F.}\ \bibnamefont
  {Marshall}}, \bibinfo {author} {\bibfnamefont {S.~H.}\ \bibnamefont
  {Risbud}}, \ and\ \bibinfo {author} {\bibfnamefont {A.}~\bibnamefont
  {Kapitulnik}},\ }\href {\doibase 10.1103/PhysRevB.88.081407} {\bibfield
  {journal} {\bibinfo  {journal} {Phys. Rev. B}\ }\textbf {\bibinfo {volume}
  {88}},\ \bibinfo {pages} {081407} (\bibinfo {year} {2013})}\BibitemShut
  {NoStop}%
\bibitem [{\citenamefont {Busch}, \citenamefont {Junod},\ and\ \citenamefont
  {Wachter}(1964)}]{EuX_absorption}%
  \BibitemOpen
  \bibfield  {author} {\bibinfo {author} {\bibfnamefont {G.}~\bibnamefont
  {Busch}}, \bibinfo {author} {\bibfnamefont {P.}~\bibnamefont {Junod}}, \ and\
  \bibinfo {author} {\bibfnamefont {P.}~\bibnamefont {Wachter}},\ }\href
  {\doibase http://dx.doi.org/10.1016/0031-9163(64)91155-2} {\bibfield
  {journal} {\bibinfo  {journal} {Physics Letters}\ }\textbf {\bibinfo {volume}
  {12}},\ \bibinfo {pages} {11 } (\bibinfo {year} {1964})}\BibitemShut
  {NoStop}%
\bibitem [{\citenamefont {Cho}(1967)}]{EuS_band_th1}%
  \BibitemOpen
  \bibfield  {author} {\bibinfo {author} {\bibfnamefont {S.~J.}\ \bibnamefont
  {Cho}},\ }\href {\doibase 10.1103/PhysRev.157.632} {\bibfield  {journal}
  {\bibinfo  {journal} {Phys. Rev.}\ }\textbf {\bibinfo {volume} {157}},\
  \bibinfo {pages} {632} (\bibinfo {year} {1967})}\BibitemShut {NoStop}%
\bibitem [{\citenamefont {M\"uller}\ and\ \citenamefont
  {Nolting}(2002)}]{EuS_band_th2}%
  \BibitemOpen
  \bibfield  {author} {\bibinfo {author} {\bibfnamefont {W.}~\bibnamefont
  {M\"uller}}\ and\ \bibinfo {author} {\bibfnamefont {W.}~\bibnamefont
  {Nolting}},\ }\href {\doibase 10.1103/PhysRevB.66.085205} {\bibfield
  {journal} {\bibinfo  {journal} {Phys. Rev. B}\ }\textbf {\bibinfo {volume}
  {66}},\ \bibinfo {pages} {085205} (\bibinfo {year} {2002})}\BibitemShut
  {NoStop}%
\bibitem [{\citenamefont {Stachow-W\'ojcik}\ \emph {et~al.}(1999)\citenamefont
  {Stachow-W\'ojcik}, \citenamefont {Story}, \citenamefont {Dobrowolski},
  \citenamefont {Arciszewska}, \citenamefont {Ga\l{}a\ifmmode~\mbox{\c{}}\else
  \c{}\fi{}zka}, \citenamefont {Kreijveld}, \citenamefont {Sw\"uste},
  \citenamefont {Swagten}, \citenamefont {de~Jonge}, \citenamefont
  {Twardowski},\ and\ \citenamefont {Sipatov}}]{EuS_PbS}%
  \BibitemOpen
  \bibfield  {author} {\bibinfo {author} {\bibfnamefont {A.}~\bibnamefont
  {Stachow-W\'ojcik}}, \bibinfo {author} {\bibfnamefont {T.}~\bibnamefont
  {Story}}, \bibinfo {author} {\bibfnamefont {W.}~\bibnamefont {Dobrowolski}},
  \bibinfo {author} {\bibfnamefont {M.}~\bibnamefont {Arciszewska}}, \bibinfo
  {author} {\bibfnamefont {R.~R.}\ \bibnamefont
  {Ga\l{}a\ifmmode~\mbox{\c{}}\else \c{}\fi{}zka}}, \bibinfo {author}
  {\bibfnamefont {M.~W.}\ \bibnamefont {Kreijveld}}, \bibinfo {author}
  {\bibfnamefont {C.~H.~W.}\ \bibnamefont {Sw\"uste}}, \bibinfo {author}
  {\bibfnamefont {H.~J.~M.}\ \bibnamefont {Swagten}}, \bibinfo {author}
  {\bibfnamefont {W.~J.~M.}\ \bibnamefont {de~Jonge}}, \bibinfo {author}
  {\bibfnamefont {A.}~\bibnamefont {Twardowski}}, \ and\ \bibinfo {author}
  {\bibfnamefont {A.~Y.}\ \bibnamefont {Sipatov}},\ }\href {\doibase
  10.1103/PhysRevB.60.15220} {\bibfield  {journal} {\bibinfo  {journal} {Phys.
  Rev. B}\ }\textbf {\bibinfo {volume} {60}},\ \bibinfo {pages} {15220}
  (\bibinfo {year} {1999})}\BibitemShut {NoStop}%
\bibitem [{\citenamefont {Keller}\ \emph {et~al.}(2002)\citenamefont {Keller},
  \citenamefont {Parker}, \citenamefont {Stankiewicz}, \citenamefont {Read},
  \citenamefont {Stampe}, \citenamefont {Kennedy}, \citenamefont {Xiong}, ,\
  and\ \citenamefont {von Molnar}}]{Keller2002}%
  \BibitemOpen
  \bibfield  {author} {\bibinfo {author} {\bibfnamefont {J.}~\bibnamefont
  {Keller}}, \bibinfo {author} {\bibfnamefont {J.}~\bibnamefont {Parker}},
  \bibinfo {author} {\bibfnamefont {J.}~\bibnamefont {Stankiewicz}}, \bibinfo
  {author} {\bibfnamefont {D.}~\bibnamefont {Read}}, \bibinfo {author}
  {\bibfnamefont {P.}~\bibnamefont {Stampe}}, \bibinfo {author} {\bibfnamefont
  {R.}~\bibnamefont {Kennedy}}, \bibinfo {author} {\bibfnamefont
  {P.}~\bibnamefont {Xiong}}, , \ and\ \bibinfo {author} {\bibfnamefont
  {S.}~\bibnamefont {von Molnar}},\ }\href {\doibase 10.1109/TMAG.2002.801977}
  {\bibfield  {journal} {\bibinfo  {journal} {IEEE Transactions on Magnetics}\
  }\textbf {\bibinfo {volume} {38}},\ \bibinfo {pages} {2673} (\bibinfo {year}
  {2002})}\BibitemShut {NoStop}%
\bibitem [{\citenamefont {Shapira}\ and\ \citenamefont
  {Reed}(1972)}]{EuS_ntype}%
  \BibitemOpen
  \bibfield  {author} {\bibinfo {author} {\bibfnamefont {Y.}~\bibnamefont
  {Shapira}}\ and\ \bibinfo {author} {\bibfnamefont {T.~B.}\ \bibnamefont
  {Reed}},\ }\href {\doibase 10.1103/PhysRevB.5.4877} {\bibfield  {journal}
  {\bibinfo  {journal} {Phys. Rev. B}\ }\textbf {\bibinfo {volume} {5}},\
  \bibinfo {pages} {4877} (\bibinfo {year} {1972})}\BibitemShut {NoStop}%
\bibitem [{\citenamefont {Moruzzi}, \citenamefont {Teaney},\ and\ \citenamefont
  {van~der Hoeven~Jr.}(1968)}]{EuS_TC_doping}%
  \BibitemOpen
  \bibfield  {author} {\bibinfo {author} {\bibfnamefont {V.}~\bibnamefont
  {Moruzzi}}, \bibinfo {author} {\bibfnamefont {D.}~\bibnamefont {Teaney}}, \
  and\ \bibinfo {author} {\bibfnamefont {B.}~\bibnamefont {van~der
  Hoeven~Jr.}},\ }\href {\doibase
  http://dx.doi.org/10.1016/0038-1098(68)90055-0} {\bibfield  {journal}
  {\bibinfo  {journal} {Solid State Communications}\ }\textbf {\bibinfo
  {volume} {6}},\ \bibinfo {pages} {461 } (\bibinfo {year} {1968})}\BibitemShut
  {NoStop}%
\bibitem [{\citenamefont {Kasuya}\ and\ \citenamefont
  {Yanase}(1968)}]{EuX_doped_transport}%
  \BibitemOpen
  \bibfield  {author} {\bibinfo {author} {\bibfnamefont {T.}~\bibnamefont
  {Kasuya}}\ and\ \bibinfo {author} {\bibfnamefont {A.}~\bibnamefont
  {Yanase}},\ }\href {\doibase 10.1103/RevModPhys.40.684} {\bibfield  {journal}
  {\bibinfo  {journal} {Rev. Mod. Phys.}\ }\textbf {\bibinfo {volume} {40}},\
  \bibinfo {pages} {684} (\bibinfo {year} {1968})}\BibitemShut {NoStop}%
\bibitem [{\citenamefont {M\"uller}, \citenamefont {Schreiber},\ and\
  \citenamefont {Schneider}(2011{\natexlab{a}})}]{EuS_MBEa}%
  \BibitemOpen
  \bibfield  {author} {\bibinfo {author} {\bibfnamefont {M.}~\bibnamefont
  {M\"uller}}, \bibinfo {author} {\bibfnamefont {R.}~\bibnamefont {Schreiber}},
  \ and\ \bibinfo {author} {\bibfnamefont {C.~M.}\ \bibnamefont {Schneider}},\
  }\href {\doibase 10.1109/TMAG.2011.2106767} {\bibfield  {journal} {\bibinfo
  {journal} {Magnetics, IEEE Transactions on}\ }\textbf {\bibinfo {volume}
  {47}},\ \bibinfo {pages} {1635} (\bibinfo {year}
  {2011}{\natexlab{a}})}\BibitemShut {NoStop}%
\bibitem [{\citenamefont {M\"uller}, \citenamefont {Schreiber},\ and\
  \citenamefont {Schneider}(2011{\natexlab{b}})}]{EuS_MBEb}%
  \BibitemOpen
  \bibfield  {author} {\bibinfo {author} {\bibfnamefont {M.}~\bibnamefont
  {M\"uller}}, \bibinfo {author} {\bibfnamefont {R.}~\bibnamefont {Schreiber}},
  \ and\ \bibinfo {author} {\bibfnamefont {C.~M.}\ \bibnamefont {Schneider}},\
  }\href {\doibase 10.1063/1.3549609} {\bibfield  {journal} {\bibinfo
  {journal} {Journal of Applied Physics}\ }\textbf {\bibinfo {volume} {109}},\
  \bibinfo {eid} {07C710} (\bibinfo {year} {2011}{\natexlab{b}})}\BibitemShut
  {NoStop}%
\bibitem [{\citenamefont {O'Mahony}\ \emph {et~al.}(2005)\citenamefont
  {O'Mahony}, \citenamefont {Smith}, \citenamefont {Budtz-Jorgensen},
  \citenamefont {Venkatesan}, \citenamefont {Lunney}, \citenamefont {McGilp},\
  and\ \citenamefont {Coey}}]{EuS_PLD1}%
  \BibitemOpen
  \bibfield  {author} {\bibinfo {author} {\bibfnamefont {D.}~\bibnamefont
  {O'Mahony}}, \bibinfo {author} {\bibfnamefont {C.}~\bibnamefont {Smith}},
  \bibinfo {author} {\bibfnamefont {C.}~\bibnamefont {Budtz-Jorgensen}},
  \bibinfo {author} {\bibfnamefont {M.}~\bibnamefont {Venkatesan}}, \bibinfo
  {author} {\bibfnamefont {J.}~\bibnamefont {Lunney}}, \bibinfo {author}
  {\bibfnamefont {J.}~\bibnamefont {McGilp}}, \ and\ \bibinfo {author}
  {\bibfnamefont {J.}~\bibnamefont {Coey}},\ }\href {\doibase
  http://dx.doi.org/10.1016/j.tsf.2005.04.081} {\bibfield  {journal} {\bibinfo
  {journal} {Thin Solid Films}\ }\textbf {\bibinfo {volume} {488}},\ \bibinfo
  {pages} {200 } (\bibinfo {year} {2005})}\BibitemShut {NoStop}%
\bibitem [{\citenamefont {Mulloy}, \citenamefont {Blau},\ and\ \citenamefont
  {Lunney}(1993)}]{EuS_PLD2}%
  \BibitemOpen
  \bibfield  {author} {\bibinfo {author} {\bibfnamefont {M.~P.}\ \bibnamefont
  {Mulloy}}, \bibinfo {author} {\bibfnamefont {W.~J.}\ \bibnamefont {Blau}}, \
  and\ \bibinfo {author} {\bibfnamefont {J.~G.}\ \bibnamefont {Lunney}},\
  }\href {\doibase 10.1063/1.352842} {\bibfield  {journal} {\bibinfo  {journal}
  {Journal of Applied Physics}\ }\textbf {\bibinfo {volume} {73}},\ \bibinfo
  {pages} {4104} (\bibinfo {year} {1993})}\BibitemShut {NoStop}%
\bibitem [{\citenamefont {Franzblau}, \citenamefont {Everett},\ and\
  \citenamefont {Lawson}(1967)}]{EuS_anisotropy}%
  \BibitemOpen
  \bibfield  {author} {\bibinfo {author} {\bibfnamefont {M.~C.}\ \bibnamefont
  {Franzblau}}, \bibinfo {author} {\bibfnamefont {G.~E.}\ \bibnamefont
  {Everett}}, \ and\ \bibinfo {author} {\bibfnamefont {A.~W.}\ \bibnamefont
  {Lawson}},\ }\href {\doibase 10.1103/PhysRev.164.716} {\bibfield  {journal}
  {\bibinfo  {journal} {Phys. Rev.}\ }\textbf {\bibinfo {volume} {164}},\
  \bibinfo {pages} {716} (\bibinfo {year} {1967})}\BibitemShut {NoStop}%
\bibitem [{\citenamefont {Tsukada}\ \emph {et~al.}(2011)\citenamefont
  {Tsukada}, \citenamefont {Luna}, \citenamefont {Hammond}, \citenamefont
  {Beasley}, \citenamefont {Zhao},\ and\ \citenamefont {Risbud}}]{Jinfeng2}%
  \BibitemOpen
  \bibfield  {author} {\bibinfo {author} {\bibfnamefont {A.}~\bibnamefont
  {Tsukada}}, \bibinfo {author} {\bibfnamefont {K.}~\bibnamefont {Luna}},
  \bibinfo {author} {\bibfnamefont {R.}~\bibnamefont {Hammond}}, \bibinfo
  {author} {\bibfnamefont {M.}~\bibnamefont {Beasley}}, \bibinfo {author}
  {\bibfnamefont {J.}~\bibnamefont {Zhao}}, \ and\ \bibinfo {author}
  {\bibfnamefont {S.}~\bibnamefont {Risbud}},\ }\href {\doibase
  10.1007/s00339-010-6136-8} {\bibfield  {journal} {\bibinfo  {journal}
  {Applied Physics A}\ }\textbf {\bibinfo {volume} {104}},\ \bibinfo {pages}
  {311} (\bibinfo {year} {2011})}\BibitemShut {NoStop}%
\bibitem [{\citenamefont {Risbud}\ and\ \citenamefont {Han}(2013)}]{Subhash1}%
  \BibitemOpen
  \bibfield  {author} {\bibinfo {author} {\bibfnamefont {S.~H.}\ \bibnamefont
  {Risbud}}\ and\ \bibinfo {author} {\bibfnamefont {Y.-H.}\ \bibnamefont
  {Han}},\ }\href {\doibase http://dx.doi.org/10.1016/j.scriptamat.2013.02.024}
  {\bibfield  {journal} {\bibinfo  {journal} {Scripta Materialia}\ }\textbf
  {\bibinfo {volume} {69}},\ \bibinfo {pages} {105 } (\bibinfo {year}
  {2013})}\BibitemShut {NoStop}%
\bibitem [{\citenamefont {Risbud}, \citenamefont {Groza},\ and\ \citenamefont
  {Kim}(1994)}]{Subhash2}%
  \BibitemOpen
  \bibfield  {author} {\bibinfo {author} {\bibfnamefont {S.~H.}\ \bibnamefont
  {Risbud}}, \bibinfo {author} {\bibfnamefont {J.~R.}\ \bibnamefont {Groza}}, \
  and\ \bibinfo {author} {\bibfnamefont {M.~J.}\ \bibnamefont {Kim}},\ }\href
  {\doibase 10.1080/01418639408240126} {\bibfield  {journal} {\bibinfo
  {journal} {Philosophical Magazine Part B}\ }\textbf {\bibinfo {volume}
  {69}},\ \bibinfo {pages} {525} (\bibinfo {year} {1994})}\BibitemShut
  {NoStop}%
\bibitem [{\citenamefont {Zhao}\ \emph {et~al.}(2009)\citenamefont {Zhao},
  \citenamefont {Holland}, \citenamefont {Unuvar},\ and\ \citenamefont
  {Munir}}]{Jinfeng1}%
  \BibitemOpen
  \bibfield  {author} {\bibinfo {author} {\bibfnamefont {J.}~\bibnamefont
  {Zhao}}, \bibinfo {author} {\bibfnamefont {T.}~\bibnamefont {Holland}},
  \bibinfo {author} {\bibfnamefont {C.}~\bibnamefont {Unuvar}}, \ and\ \bibinfo
  {author} {\bibfnamefont {Z.~A.}\ \bibnamefont {Munir}},\ }\href {\doibase
  http://dx.doi.org/10.1016/j.ijrmhm.2008.06.004} {\bibfield  {journal}
  {\bibinfo  {journal} {International Journal of Refractory Metals and Hard
  Materials}\ }\textbf {\bibinfo {volume} {27}},\ \bibinfo {pages} {130 }
  (\bibinfo {year} {2009})}\BibitemShut {NoStop}%
\bibitem [{\citenamefont {Van~der Pauw}(1958)}]{VdP1958}%
  \BibitemOpen
  \bibfield  {author} {\bibinfo {author} {\bibfnamefont {L.}~\bibnamefont
  {Van~der Pauw}},\ }\href@noop {} {\bibfield  {journal} {\bibinfo  {journal}
  {Philips Tech. Rev.}\ }\textbf {\bibinfo {volume} {20}},\ \bibinfo {pages}
  {220} (\bibinfo {year} {1958})}\BibitemShut {NoStop}%
\bibitem [{\citenamefont {Dolan}\ and\ \citenamefont
  {Osheroff}(1979)}]{2D_conduction}%
  \BibitemOpen
  \bibfield  {author} {\bibinfo {author} {\bibfnamefont {G.~J.}\ \bibnamefont
  {Dolan}}\ and\ \bibinfo {author} {\bibfnamefont {D.~D.}\ \bibnamefont
  {Osheroff}},\ }\href {\doibase 10.1103/PhysRevLett.43.721} {\bibfield
  {journal} {\bibinfo  {journal} {Phys. Rev. Lett.}\ }\textbf {\bibinfo
  {volume} {43}},\ \bibinfo {pages} {721} (\bibinfo {year} {1979})}\BibitemShut
  {NoStop}%
\bibitem [{\citenamefont {McGuire}\ and\ \citenamefont
  {Shafer}(1964)}]{Eu_mag_compounds}%
  \BibitemOpen
  \bibfield  {author} {\bibinfo {author} {\bibfnamefont {T.~R.}\ \bibnamefont
  {McGuire}}\ and\ \bibinfo {author} {\bibfnamefont {M.~W.}\ \bibnamefont
  {Shafer}},\ }\href {\doibase 10.1063/1.1713568} {\bibfield  {journal}
  {\bibinfo  {journal} {Journal of Applied Physics}\ }\textbf {\bibinfo
  {volume} {35}},\ \bibinfo {pages} {984} (\bibinfo {year} {1964})}\BibitemShut
  {NoStop}%
\bibitem [{\citenamefont {Schiller}\ and\ \citenamefont
  {Nolting}(1999)}]{TC_thickness1}%
  \BibitemOpen
  \bibfield  {author} {\bibinfo {author} {\bibfnamefont {R.}~\bibnamefont
  {Schiller}}\ and\ \bibinfo {author} {\bibfnamefont {W.}~\bibnamefont
  {Nolting}},\ }\href {\doibase
  http://dx.doi.org/10.1016/S0038-1098(98)00593-6} {\bibfield  {journal}
  {\bibinfo  {journal} {Solid State Communications}\ }\textbf {\bibinfo
  {volume} {110}},\ \bibinfo {pages} {121 } (\bibinfo {year}
  {1999})}\BibitemShut {NoStop}%
\bibitem [{\citenamefont {Huang}\ \emph {et~al.}(1993)\citenamefont {Huang},
  \citenamefont {Mankey}, \citenamefont {Kief},\ and\ \citenamefont
  {Willis}}]{TC_thickness2}%
  \BibitemOpen
  \bibfield  {author} {\bibinfo {author} {\bibfnamefont {F.}~\bibnamefont
  {Huang}}, \bibinfo {author} {\bibfnamefont {G.~J.}\ \bibnamefont {Mankey}},
  \bibinfo {author} {\bibfnamefont {M.~T.}\ \bibnamefont {Kief}}, \ and\
  \bibinfo {author} {\bibfnamefont {R.~F.}\ \bibnamefont {Willis}},\ }\href
  {\doibase 10.1063/1.352477} {\bibfield  {journal} {\bibinfo  {journal}
  {Journal of Applied Physics}\ }\textbf {\bibinfo {volume} {73}},\ \bibinfo
  {pages} {6760} (\bibinfo {year} {1993})}\BibitemShut {NoStop}%
\bibitem [{\citenamefont {Tsui}\ \emph {et~al.}(2000)\citenamefont {Tsui},
  \citenamefont {Smoak}, \citenamefont {Nath},\ and\ \citenamefont
  {Eom}}]{TC_strain}%
  \BibitemOpen
  \bibfield  {author} {\bibinfo {author} {\bibfnamefont {F.}~\bibnamefont
  {Tsui}}, \bibinfo {author} {\bibfnamefont {M.~C.}\ \bibnamefont {Smoak}},
  \bibinfo {author} {\bibfnamefont {T.~K.}\ \bibnamefont {Nath}}, \ and\
  \bibinfo {author} {\bibfnamefont {C.~B.}\ \bibnamefont {Eom}},\ }\href
  {\doibase 10.1063/1.126363} {\bibfield  {journal} {\bibinfo  {journal}
  {Applied Physics Letters}\ }\textbf {\bibinfo {volume} {76}},\ \bibinfo
  {pages} {2421} (\bibinfo {year} {2000})}\BibitemShut {NoStop}%
\bibitem [{\citenamefont {Fried}, \citenamefont {Fejer},\ and\ \citenamefont
  {Kapitulnik}()}]{futureSagnac}%
  \BibitemOpen
  \bibfield  {author} {\bibinfo {author} {\bibfnamefont {A.~D.}\ \bibnamefont
  {Fried}}, \bibinfo {author} {\bibfnamefont {M.~M.}\ \bibnamefont {Fejer}}, \
  and\ \bibinfo {author} {\bibfnamefont {A.}~\bibnamefont {Kapitulnik}},\
  }\href@noop {} {\enquote {\bibinfo {title} {A scanning, all-fiber sagnac
  interferometer for high resolution magneto-optic measurements at
  $820\mathrm{nm}$},}\ }\bibinfo {note} {To be published}\BibitemShut {NoStop}%
\bibitem [{\citenamefont {Xia}\ \emph {et~al.}(2006)\citenamefont {Xia},
  \citenamefont {Beyersdorf}, \citenamefont {Fejer},\ and\ \citenamefont
  {Kapitulnik}}]{xia:062508}%
  \BibitemOpen
  \bibfield  {author} {\bibinfo {author} {\bibfnamefont {J.}~\bibnamefont
  {Xia}}, \bibinfo {author} {\bibfnamefont {P.~T.}\ \bibnamefont {Beyersdorf}},
  \bibinfo {author} {\bibfnamefont {M.~M.}\ \bibnamefont {Fejer}}, \ and\
  \bibinfo {author} {\bibfnamefont {A.}~\bibnamefont {Kapitulnik}},\ }\href
  {\doibase 10.1063/1.2336620} {\bibfield  {journal} {\bibinfo  {journal}
  {Applied Physics Letters}\ }\textbf {\bibinfo {volume} {89}},\ \bibinfo {eid}
  {062508} (\bibinfo {year} {2006})}\BibitemShut {NoStop}%
\bibitem [{\citenamefont {Cheung}, \citenamefont {Geohegan},\ and\
  \citenamefont {Chen}(1994)}]{PLD_book}%
  \BibitemOpen
  \bibfield  {author} {\bibinfo {author} {\bibfnamefont {J.~T.}\ \bibnamefont
  {Cheung}}, \bibinfo {author} {\bibfnamefont {D.~B.}\ \bibnamefont
  {Geohegan}}, \ and\ \bibinfo {author} {\bibfnamefont {L.-C.}\ \bibnamefont
  {Chen}},\ }\href@noop {} {\emph {\bibinfo {title} {Pulsed Laser Deposition of
  Thin Films}}},\ edited by\ \bibinfo {editor} {\bibfnamefont {D.~B.}\
  \bibnamefont {Chrisey}}\ and\ \bibinfo {editor} {\bibfnamefont {G.~K.}\
  \bibnamefont {Hubler}}\ (\bibinfo  {publisher} {Wiley},\ \bibinfo {year}
  {1994})\ Chap.\ \bibinfo {chapter} {1, 5, 6}\BibitemShut {NoStop}%
\bibitem [{\citenamefont {Okamoto}(1990)}]{EuS_phase_diagram}%
  \BibitemOpen
  \bibfield  {author} {\bibinfo {author} {\bibfnamefont {H.}~\bibnamefont
  {Okamoto}},\ }\enquote {\bibinfo {title} {{Eu-S Phase Diagram}},}\ in\ \href
  {http://www1.asminternational.org/asmenterprise/apd/ViewAPD.aspx?id=901019}
  {\emph {\bibinfo {booktitle} {Alloy phase diagrams center}}}\ (\bibinfo
  {publisher} {ASM International},\ \bibinfo {year} {1990})\BibitemShut
  {NoStop}%
\bibitem [{\citenamefont {Yaws}(2011)}]{elements}%
  \BibitemOpen
  \bibfield  {author} {\bibinfo {author} {\bibfnamefont {C.~L.}\ \bibnamefont
  {Yaws}},\ }\href@noop {} {\emph {\bibinfo {title} {Yaws' Handbook of
  Properties of the Chemical Elements}}}\ (\bibinfo  {publisher} {Knovel},\
  \bibinfo {year} {2011})\BibitemShut {NoStop}%
\bibitem [{\citenamefont {Metev}\ and\ \citenamefont
  {Sendova}(1988)}]{Metev1988}%
  \BibitemOpen
  \bibfield  {author} {\bibinfo {author} {\bibfnamefont {S.}~\bibnamefont
  {Metev}}\ and\ \bibinfo {author} {\bibfnamefont {M.}~\bibnamefont
  {Sendova}},\ }\enquote {\bibinfo {title} {Thin-film compounds formation with
  pulsed laser-plasma fluxes},}\ in\ \href@noop {} {\emph {\bibinfo {booktitle}
  {Third International Conference on Trends in Quantum Electronics}}}\
  (\bibinfo  {publisher} {European Physical Society},\ \bibinfo {year} {1988})\
  p.\ \bibinfo {pages} {260}\BibitemShut {NoStop}%
\bibitem [{\citenamefont {Galassi}\ \emph {et~al.}(2009)\citenamefont
  {Galassi}, \citenamefont {Davies}, \citenamefont {Theiler}, \citenamefont
  {Gough}, \citenamefont {Jungman}, \citenamefont {Alken}, \citenamefont
  {Booth},\ and\ \citenamefont {Rossi}}]{gsl}%
  \BibitemOpen
  \bibfield  {author} {\bibinfo {author} {\bibfnamefont {M.}~\bibnamefont
  {Galassi}}, \bibinfo {author} {\bibfnamefont {J.}~\bibnamefont {Davies}},
  \bibinfo {author} {\bibfnamefont {J.}~\bibnamefont {Theiler}}, \bibinfo
  {author} {\bibfnamefont {B.}~\bibnamefont {Gough}}, \bibinfo {author}
  {\bibfnamefont {G.}~\bibnamefont {Jungman}}, \bibinfo {author} {\bibfnamefont
  {P.}~\bibnamefont {Alken}}, \bibinfo {author} {\bibfnamefont
  {M.}~\bibnamefont {Booth}}, \ and\ \bibinfo {author} {\bibfnamefont
  {F.}~\bibnamefont {Rossi}},\ }\href {http://www.gnu.org/software/gsl/} {\emph
  {\bibinfo {title} {GNU Scientific Library: Reference Manual}}}\ (\bibinfo
  {publisher} {Network Theory Limited},\ \bibinfo {year} {2009})\BibitemShut
  {NoStop}%
\bibitem [{\citenamefont {Williams}, \citenamefont {Kelley}\ \emph
  {et~al.}()\citenamefont {Williams}, \citenamefont {Kelley} \emph
  {et~al.}}]{gnuplot}%
  \BibitemOpen
  \bibfield  {author} {\bibinfo {author} {\bibfnamefont {T.}~\bibnamefont
  {Williams}}, \bibinfo {author} {\bibfnamefont {C.}~\bibnamefont {Kelley}},
  \emph {et~al.},\ }\href {http://www.gnuplot.info/} {\enquote {\bibinfo
  {title} {gnuplot: An interactive plotting program},}\ }\BibitemShut {NoStop}%
\end{thebibliography}%

\end{document}